\documentclass[%
 reprint,
 amsmath,amssymb,
 aps,
]{revtex4-2}
\usepackage{graphicx}
\usepackage{dcolumn}
\usepackage[normalem]{ulem}
\usepackage{bm}

\begin{document}



\title{Tailored Hotspots from Airy-Based Surface Plasmon Polaritons}

\author{Rosario Martínez-Herrero}
\email{r.m-h@fis.ucm.es}
\affiliation{Department of Optics, Faculty of Physical Sciences, University Complutense de Madrid, Pza.\ Ciencias 1, 28040 Madrid, Spain}

\author{\'Angel S. Sanz}
\email{a.s.sanz@fis.ucm.es}
\affiliation{Department of Optics, Faculty of Physical Sciences, University Complutense de Madrid, Pza.\ Ciencias 1, 28040 Madrid, Spain}

\author{Javier Hernandez-Rueda}
\email{fj.hernandez.rueda@ucm.es}
\affiliation{Department of Optics, Faculty of Physical Sciences, University Complutense de Madrid, Pza.\ Ciencias 1, 28040 Madrid, Spain}

\date{\today}

\begin{abstract}

Surface plasmons have attracted growing interest from the photonics community due to their inherent ability to controllably confine light below the diffraction limit and their direct application in trapping and transporting matter at the nanoscale. This method, known as plasmonic tweezers, employs confined fields generated by either localized plasmons or surface plasmon polaritons (SPP), which originate in the vicinity of nanostructure-based traps or across structureless platforms, respectively. Herein, we present a new theoretical method for generating intense light hotspots and engineering their features by overlapping Airy SPPs (ASPP) at a smooth dielectric-metal interface. We coherently add pairs of Hermite-Gauss modes that belong to a novel complete basis set of finite-energy ASPPs, which yield highly confined plasmonic hotspots ($\approx \lambda$/10) without the need of using any nanostructured platform. Mode order and relative spacing parameters can be used to tailor the intensity and quality factor of said hotspots, largely outperforming their Gaussian-only-based counterparts. Our method opens a promising venue to confine light at the nanoscale using ASPP-based structured light, which helps to advance the development of structureless plasmonic tweezers and holds promising potential for its application in optical signal processing and plasmonic circuitry.


\end{abstract}


\maketitle

\newpage


\textit{Introduction—}Over the past decades, surface plasmons (SPs) have gathered the attention of the nanophotonics community due to their inherently strong light-matter interaction at sub-wavelength scales and their potential for novel applications \cite{maradudin:PhysRep:2005,doi:10.1021/nl802044t,Manjavacas:09,Zhang:JPhysD:2012,xiangang:IEEEPhotJ:2012,lin:PRL:2012,puerto2016femtosecond,anwar:DigCommunNet:2018}. SPs consist in electromagnetic waves coupled to electron oscillations on metals. These waves originate in close proximity to metallic nanostructures or nanoparticles leading to localized surface plasmons (LPs) or propagate across smooth dielectric-metal interfaces as surface plasmon polaritons (SPPs). Both types of SPs originate highly localized intense light fields with sizes well below the diffraction limit, which make them particularly attractive to detect, manipulate and enhance light-matter interactions at the nanoscale. For instance, these unique properties have contributed to develop detection-based applications ranging from subwavelength imaging \cite{zayats2005nano,minovich2011generation,rotenberg2014mapping} to biosensing \cite{mejia2018plasmonic,rodrigo2015mid,brolo2012plasmonics} and to enhanced spectroscopic methods \cite{camden2008controlled,jahn2016plasmonic}. SP-assisted enhanced fields have also been exploited to manipulate light in a variety of carefully designed nanophotonic devices, e.g., merged plasmonic nano-structures with 2D materials \cite{miroshnichenko2013polarization,gong2018nanoscale,bliokh2017optical,noordam2020plasmon}. Moreover, smart combinations of state-of-the-art structured light methods with nanofabricated plasmonic systems provide new opportunities for applications in quantum technologies, advanced microscopy methods, and plasmonic circuitry \cite{forbes:NatPhotonics:2021,dholakia:NatPhoton:2008,dholakia:NatMethods:2014,fang:LSA:2015}.


More recently, the ability of SPs to originate highly localized intense fields has been exploited for trapping and manipulating nanoparticles, biomolecules and quantum dots at the nanoscale \cite{liu2005focusing,min2013focused,zhang2021plasmonic,dreher2023focused}. Plasmonic-based tweezers leverage the unique propagation features of SPs, allowing for the precise control of nano-objects with sizes below the diffraction limit. Typically, plasmonic tweezers employ LPs originated in the vicinity of machined metallic nanostructures \cite{christodoulides:OL:2010,li:PRL:2011,Klein:OptLett:2012,minovich:LasPhotRev:2014}. In contrast, structureless plasmonic tweezers exploit the intrinsic properties of tailored SPPs launched across bare dielectric-metal interfaces, which conveniently avoid the meticulous procedures needed to fabricate the nanostructures that aid to generate LPs. The formal description of SPPs can be done using many-body considerations focused on the electronic response of solids \cite{echenique:RepProgPhys:2006}, although theoretical descriptions based on Maxwell’s equations are more commonly applied and usually suffice to cover most application-oriented scenarios. However, while versatile, these methods do not provide the transversal structure of SPPs that is crucial to develop novel trapping methods. Therefore, a full description of the propagation features and confinement of SPPs entails a theoretical method beyond the single plane-wave approximation \cite{manjavacas:PRA:2016,hu:ResPhys:2021,ruan:OL:2024,chen:OL:2023}. In a recent work, we introduced a novel methodology to describe SPPs based on a Hermite-Gaussian mode expansion of their electromagnetic field \cite{hernandez2024engineering}. Our proposed description was demonstrated to be general and can be used to engineer the propagation properties of any self-bending SPP and the generation of hotspots. This approach is particularly useful for tailoring the properties of intense structured light at the nanoscale and holds potential for the development of structureless plasmonic tweezers and for new applications in plasmonic circuitry.


In this Letter, we apply our aforementioned theoretical methodology to generate and to manipulate nano-sized hotspots through the coherent superposition of Airy surface plasmon polaritons (ASPPs). Our method is based on a Hermite-Gaussian mode expansion of the electromagnetic fields to describe finite-energy ASPPs under paraxial propagation conditions along a metal-dielectric interface. This approach allows for precise control over the intensity, spatial position, and size of the hotspots generated through the degenerate superposition of multiple ASPPs (i.e., same order $n$). We demonstrate this by analyzing the interaction of overlapping ASPPs at a silver-air interface, revealing enhanced hotspot intensities and tunable spatial features via the $n$ mode selection and the $x_0$ spacing parameter. 


\textit{ASPP-based light hotspots—}In the following, we illustrate how to combine specific ASPP-modes to generate hotspots with enhanced intensities and quality factors that are superior to those generated by their Gaussian-only-based counterparts. We combine pairs of modes of a basis set of finite-energy self-bending SPPs beams at a metal-dielectric interface in the paraxial approximation. Recently, we reported on the use of such complete basis set to investigate the propagation of single mode Airy surface plasmon polaritons \cite{hernandez2024engineering}. Accelerating finite-energy Airy beams were experimentally demonstrated elsewhere by imposing a cubic-phase modulation to the spatial phase by means of a spatial light modulator \cite{christodoulides:OptLett:2007}. Herein, we briefly describe and apply our newly developed basis set to readily generate and control the features of hotspots across a silver-glass interface. 

The spatial location of the hotspots can be controlled by simply tuning the relative distance between the ASPP modes that overlap, which we define as 2$x_0$. Critically, the mode order $n$ choice is crucial and conditions the location coordinates of the hotspot, as we shall discuss later in the manuscript. We consider the propagation of ASPPs at a metal-dielectric interface ($x-z$ plane), where the dielectric corresponds to $y > 0$ and the metal to $y < 0$. The dielectric and metal permittivities are denoted as $\varepsilon_d$ and $\varepsilon_m = \varepsilon'_m + i \varepsilon''_m$, respectively, where $\varepsilon_d + \varepsilon'_m < 0$. The complex wave number of the ASPP can then be defined as

\begin{equation}
 k_{\rm sp} = k_0 \sqrt{\frac{\varepsilon_m \varepsilon_d}{\varepsilon_m + \varepsilon_d}} ,
 \label{eq1}
\end{equation}

\noindent where $k_0 =2\pi/\lambda$, with a wavelength $\lambda = 633$\,nm. Our current study case considers an air-silver interface; therefore, we employ the dielectric function $\varepsilon_m = -18.3 + 0.5 i$ at 633\,nm \cite{johnson:PRB:1972}. As mentioned above, tailored ASPP-based hotspots can be achieved by overlapping several ASPP basis elements. Here, we consider the simplest case of a coherent symmetric superposition of two elements of the basis of finite-energy ASPPs for a given $n$ as 
\begin{equation}
 G_{\rm sn}(x,z) = G_{\rm Ln}(x,z) + G_{\rm Rn}(x,z) ,
 \label{eq2}
\end{equation}
where 
\begin{eqnarray}
  G_{\rm Ln} & = & \int_{-\infty}^\infty e^{\frac{iu^3}{3}} \Psi_n(u) e^{i|k_{\rm sp}| u(x+x_0)} e^{-ik^*_{\rm sp} u^2 z/2} du ,
 \label{eq3} \\
  G_{\rm Rn} & = & \int_{-\infty}^\infty e^{\frac{iu^3}{3}} \Psi_n(u) e^{i|k_{\rm sp}| u(x-x_0)} e^{-ik^*_{\rm sp} u^2 z/2} du.
 \label{eq4}
\end{eqnarray}

\noindent $\Psi_n(u)$ defines a complete basis set of Hermite-Gaussian modes given by

\begin{figure}[h!]
\centering
\includegraphics[width=1\linewidth]{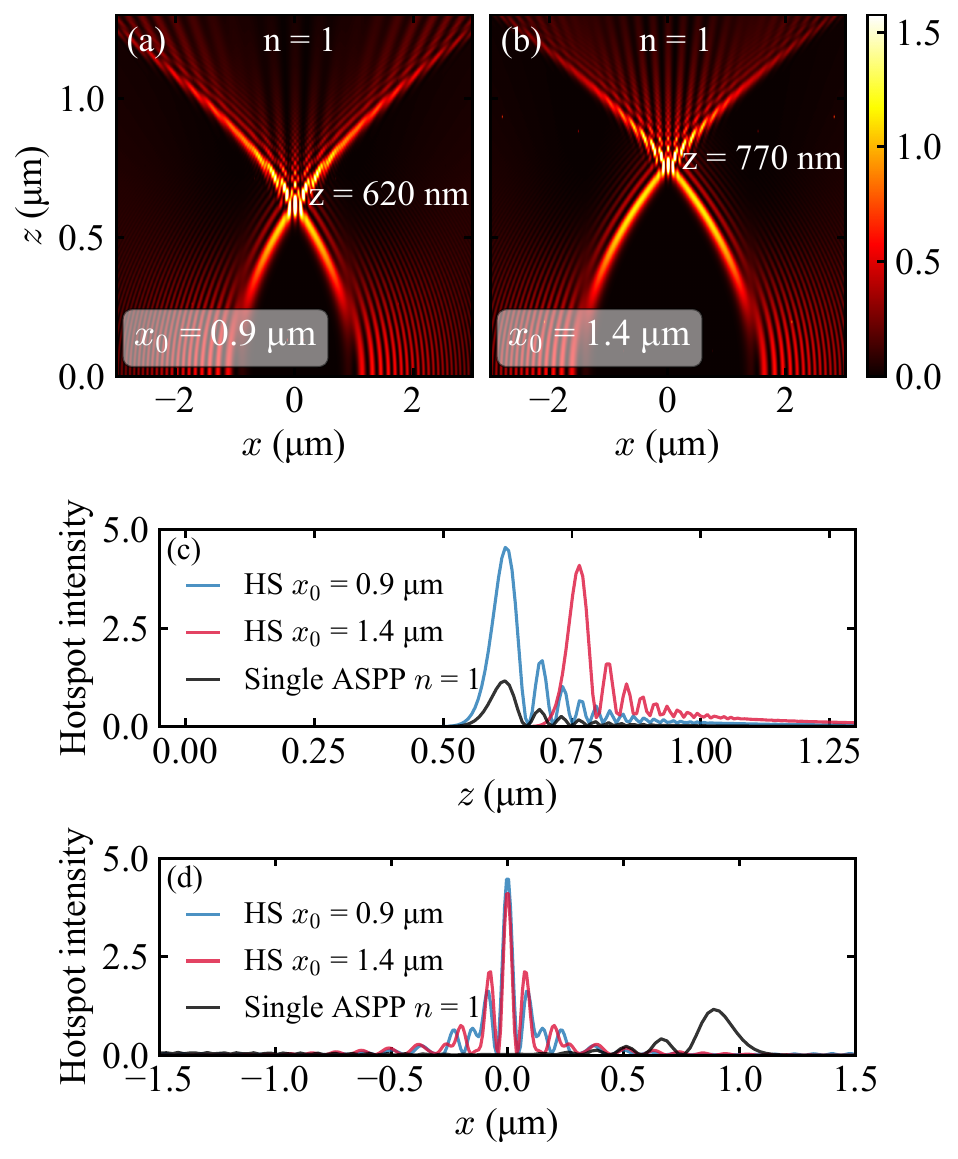}
\caption {(a)-(b) Density plots of the intensity of two overlapping ASPPs at a silver-air interface with $n=1$ and inter-beam separations equal to (a) $x_0=0.9$~$\mu$m and (b) $x_0=1.4$~$\mu$m.
Plots of the intensity cross-sections that intersect the hotspot in map (a) along the $z$ propagation direction (c) and the $x$ transverse direction (d). The black, blue and red cross sections presented in (d) are extracted at $z_{black}=620$~nm, $z_{blue}=620$~nm and $z_{red}=770$~nm. }
\label{fig1}
\end{figure}

\begin{eqnarray}
  \Psi_n(u) =a_nH_n(\alpha u) e^{-\alpha^2 u^2 /2},
 \label{eq5}
\end{eqnarray}

\noindent where $\alpha$ is a dimensionless constant, $H_n$ is an nth-order Hermite polynomial, and $a_n$ can be expressed as

\begin{eqnarray}
 a_n^2 = \frac{ \alpha}{2^n n! \sqrt{\pi}}.  
  \label{eq6}
\end{eqnarray}

More details on the development of the basis of the finite energy ASPPs can be found in the Appendix \cite{hernandez2024engineering}. For a given $n$, Eq.\,\ref{eq1} provides a symmetric superposition of two ASPPs whose first lobe at $z=0$ is separated by a distance of $2x_0$. Therefore, the parameter $x_0$ dictates the intensity of the hotspots and their location along the propagation direction $z$ with $x =$ 0. The intensity of the overlap renders

\begin{eqnarray}
 I_{\rm sn}(x,z) =&& |G_{\rm Ln}(x,z)|^2 + |G_{\rm Rn}(x,z)|^2+\nonumber\\
&&
 2 {\rm Re} \left[ G_{\rm Ln}^*(x,z) G_{\rm Rn}(x,z) \right] .
 \label{eq7}
\end{eqnarray}

\noindent Figs.~\ref{fig1}(a)-(b) display density plots of symmetric ASPP superpositions for $n=1$. These maps illustrate the generation of hotspots for two relative beam separations of $2x_0^a=1.8$~$\mu$m and $2x_0^b=2.8$~$\mu$m, respectively. As the beam pairs $G_{\rm Ln}$ and $G_{\rm Rn}$ propagate along $z$, they symmetrically approach each other along the transverse coordinate with respect to $x=0$. Their overlap for these specific values $x_0$ results in the generation of sharply localized hotspots at $z_{hs}^a=625$\,nm and $z_{hs}^b=775$\,nm with $x_{hs}=0$. These qualitatively illustrate how the separation $x_0$ between ASPPs, as well as their inherent curvature, dictate the position $z_{hs}$ of the hotspot. We will quantitatively investigate the $z_{hs}$-$x_0$ relationship later in the text. Our specific $2x_0$ separation choice in Fig.~\ref{fig1}(a) provides the most intense hotspot of all $n=1$ superposition cases $I_{\rm s1}(x,z)$. Therein, we obtain a hotspot intensity that is a factor 4.5 times higher than the highest intensity individually achieved by its ASPPs components $|G_{\rm Ln}(x,z)|^2$ and $|G_{\rm Rn}(x,z)|^2$. A simple coordinate transformation along the transverse axis $x$ or a shift introduced to one of the beams through an additional $z_0$ can help break the mirror symmetry of the superposition defined in Eq.~\ref{eq7}. Consequently, such symmetry breaking can be used to control the lateral position $x_{hs}$ of the hotspots \cite{minovich2011generation}. In Figs.~\ref{fig1}(c) and (d), the hotspot intensity can be quantitatively compared to the maximum intensity achieved by a single ASPP with $n=1$. We show the intensity profiles that intersect the hotspots presented in panels (a) and (b) along the propagation $z$ and transverse $x$ directions. The cross sections in blue and red illustrate narrower intensity distributions, along $x$ and $z$, and superior hotspot maximum intensities with respect to those reached by a single ASPP. As mentioned before, the maximum enhancement $I^{max.}_{hs}/I^{max.}_{single}$ reaches a factor 4.5, for the case presented in panel (a), and a factor 4.1 for the case displayed in panel (b). In addition, hotspots originate narrower distributions than single ASPPs as shown in Figs.~\ref{fig1}(c)-(d), which in the present examples leads to widths (in terms of the full-width at half-maximum, FWHM) of $d_{a}^x=48$\,nm and $d_{a}^z=55$\,nm, and $d_{b}^x=40$\,nm and $d_{b}^z=45$\,nm. Consequently, narrower hotspot distributions will allow us to engineer higher quality factors. These results exemplify how ASPP-based hotspots with large intensities can be readily generated for different $z$ coordinates.





\begin{figure}[!t]
\centering
\includegraphics[width=\linewidth]{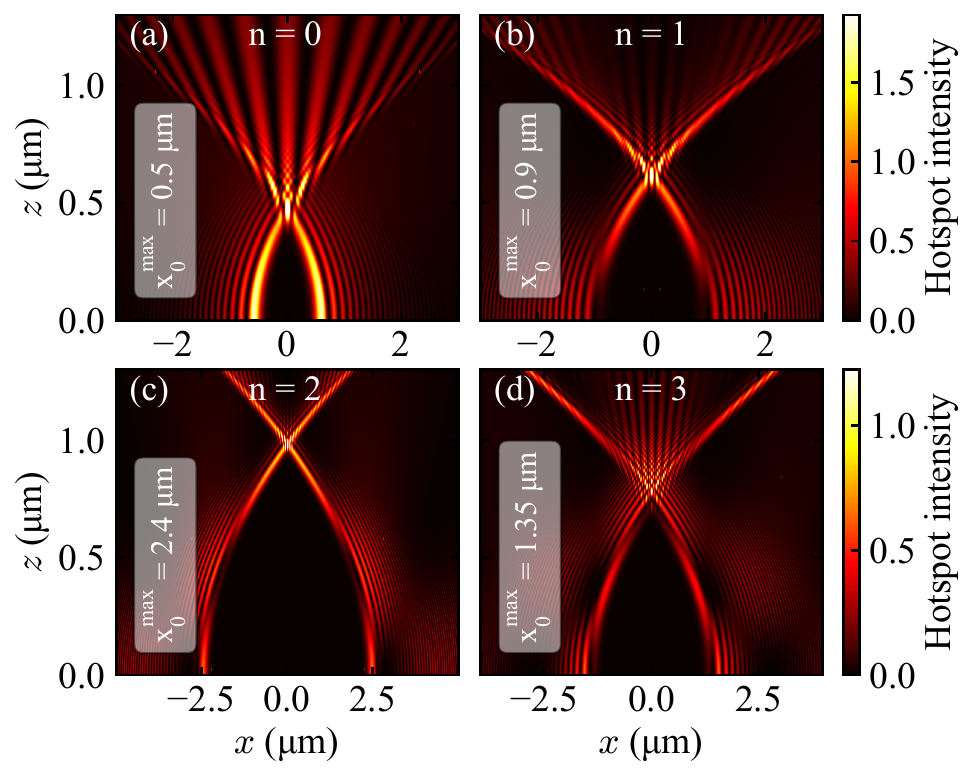}
\caption{(a)-(d) Density plots of the intensity of two overlapping APPS at a silver-air interface as a function the propagation $z$ and transverse $x$ directions for $n$=0-3, respectively. The $x_0$ separations were chosen to provide the most intense hotspots for each order $n$.}
\label{fig2}
\end{figure}

\textit{Hotspot engineering using ASPP order—}Figures.~\ref{fig2}(a)-(d) present density plots of symmetric intensity distributions of overlapped ASPPs for different modes of our Hermite-Gaussian basis set of self-bending SPPs \cite{hernandez2024engineering}. We have selected specific $x_0$ values for which the so-generated hotspots provide the largest intensity enhancement of each mode superposition $I_{\rm sn}(x,z)$, with $n = 0,1,2,3$. Larger ASPP mode separations $x_0$ aid to generate hotspots at larger $z_{hs}$ locations. Overall, the maximum intensity of single ASPP modes increases from $z=$ 0.4\,$\mu$m to 1.3\,$\mu$m as $n$ decreases. In contrast, all $I_{\rm sn}(x,z)$ superpositions provide hotspot intensities ranging from 2.2 up to 4.7, which normalized to the maxima generated by their single-mode ASPP counterparts yield enhancement factors $I^{max.}_{hs}/I^{max.}_{single}$ well above 3. In addition, our results reveal sharper hotspots for increasing $n$ values. For instance, the $n=0$ and $n=2$ cases provide hotspots widths (FWHM) of $d_{0}^x=72$\,nm and $d_{2}^x=32$\,nm, respectively. Crucially, this indicates that using ASPP superposition parameters one can control the quality factor of hotspots that lead to enhanced light intensities. Tailoring hotspots with sub-50-nm spatial features holds great potential to contribute to advance structureless trapping methods to confine sub-10-nm particles. Confining methods for such small particles typically rely on metamaterial-assisted tweezers to trap for example biomolecules and dielectric nanobeads \cite{kotsifaki2020fano,kotsifaki2021dynamic,kotsifaki2024hybrid}. Other $x_0$-$q$ combinations will provide hotspots with a variety of enhancement factors at different $z_{hs}$-$x_{hs}$ locations as we will discuss in the following. 

\begin{figure}[!t]
\centering
\includegraphics[width=\linewidth]{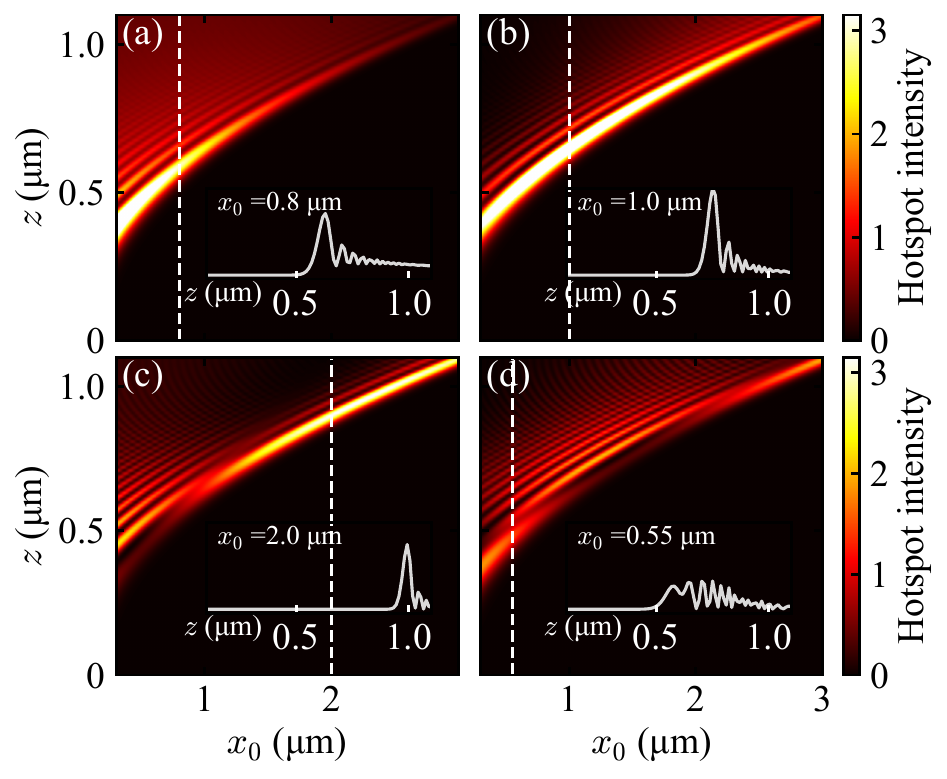}
\caption{Density plots of the intensity of hotspots as a function of the propagation direction $z$ and the separation $x_0$ between two overlapping APPS at a silver-air interface. The (a)-(d) panels illustrate maps corresponding to $n$=0-3, respectively. The insets display $I_n$-$z$ profiles for separations $x_0$ at which the so-generated hotspots have a maximum intensity. }
\label{fig3}
\end{figure}

To provide a complete overview of the features of hotspots and aid their visualization, Figs.~\ref{fig3}(a)-(d) present density plots with hotspot intensities as a function of $x_0$ and $z$ for $n = 0,1,2,3$. Note here that we only consider hotspots generated with separations $x_0$ that are at least two times the FWHM of the first lobe of the individual Airy beams at $x=0$ and $n=0$ (i.e., $x_0>$300\,nm). Since we consider symmetric distributions, the maps compile the hotspot intensity profiles along $z$ at $x=0$ where the hotspots are located. The insets illustrate examples of such profiles for relevant $x_0$ values that result in a high intensity (see the color code). At first glance, the maps reveal that higher $n$-order superpositions yield weaker intensities. Naturally, the hotspot maxima as a function of $x_0$ slide along the parabolic shape of their individual ASPP components, which is where the components intersect and originate the hotspots. As a consequence, superpositions $I_{\rm sn}(x,z)$ with increasing $n$ values can be used to generate intense hotspots at ever-larger $z_{hs}$ locations when setting longer $x_0$ separations. This feature has potential for trapping applications as it offers the versatility of generating hotspots with considerably large intensities at long $z$ propagation distances by exploiting the superposition of higher-order modes. 


\begin{figure}[!t]
\centering
\includegraphics[width=0.8\linewidth]{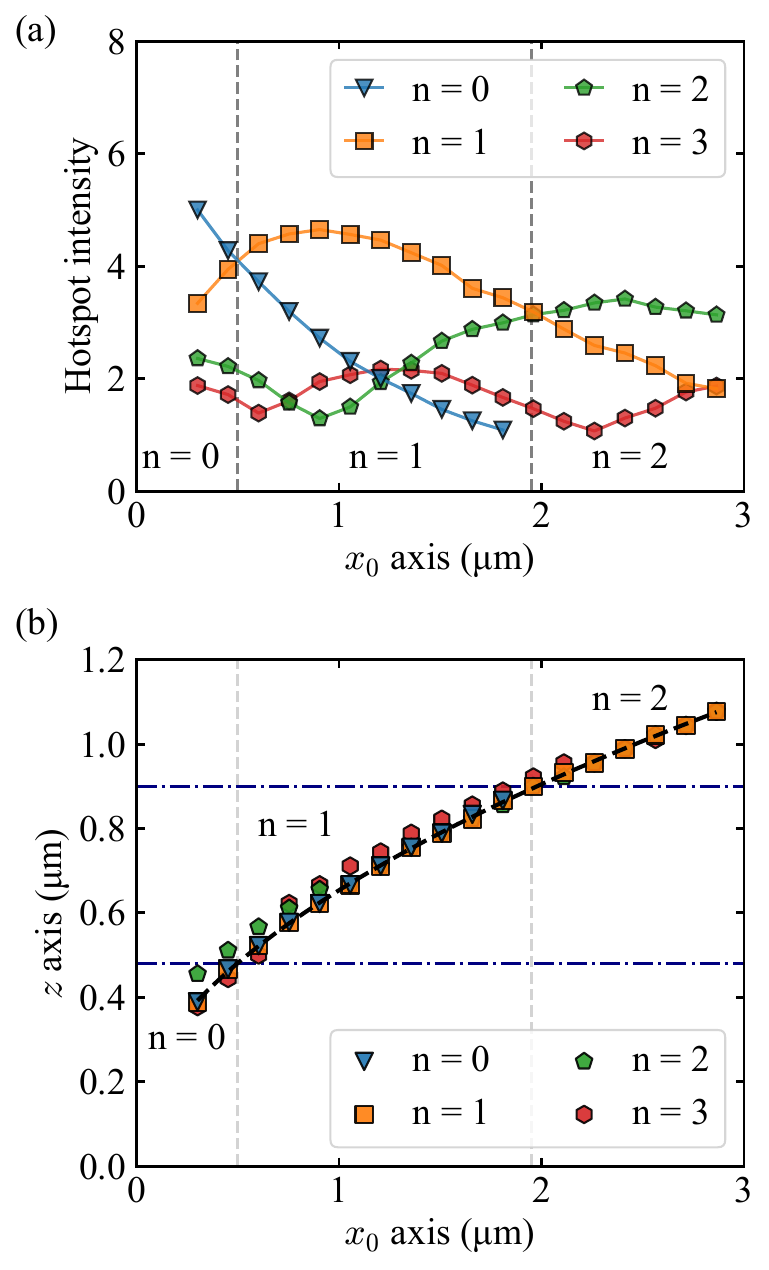}
\caption{ Graphs of the (a) hotspot intensity and (b) position along the propagation direction $z$ as a function of the separation $x_0$ between two overlapping APPS for $n$=0-3. The dashed black line is a fit to the data. }
\label{fig4}
\end{figure}

In Figures.~\ref{fig4}(a) and (b), we plot the maximum hotspot intensity $I_{\rm sn}^{max}(x,z)$ and their $z_{hs}$ location as a function of $x_0$, respectively. These data are extracted from the numerical calculations shown in Fig.~\ref{fig3}. The intensity of the hotspots as a function of $x_0$ evolves differently for different $n$ values (see different markers), which is consistent with the propagation characteristics of the individual components, $G_{\rm Ln}(x,z)$ and $G_{\rm Rn}(x,z)$, which form each superposition. The case $n=0$ illustrates a decreasing trend whereas the case $n=1$ shows an increase in intensity that peaks at a separation $x_0 =$ 0.9\,$\mu$m. The cases $n=2$ and $3$ present a more complex intensity dependency with $x_0$, starting with a decreasing behavior similar to the case $n=0$ and then increasing again to reach maxima at 2.4\,$\mu$m (pentagons) and 1.3\,$\mu$m (hexagons), as anticipated in Figs.~\ref{fig2}(c) and (d), respectively. The vertical gray lines at $x_0=$ 0.50\,$\mu$m and 1.95\,$\mu$m mark the boundaries of the beam separation regimes, $x_0$, for which specific values of $n$ lead to more intense hotspots, that is, $n=0,1,2$ for the first, second, and third $x_0$ intervals, respectively. Fig.~\ref{fig4}(b) quantitatively illustrates how the location of the hotspots $z_{hs}$ monotonously grows with increasing values of $x_0$. The $z_{hs}$-$x_0$ relationship can be predicted by introducing the function $z_{hs}=z_0\sqrt{\gamma x_0}$. The fit of this function to the simulated data is shown in Fig.~\ref{fig4}(b) as a dashed-black line and yields  $z_0=0.96$\,$\mu$m and $\gamma=0.64$\,$\mu$m$^{-1}$. In panel (b), the intersections of the vertical gray lines of panel (a) with the simulated $z_{hs}$ data provide the boundaries of the $z$ regions, marked by horizontal blue lines, where the most intense hotspots can be generated using the indicated $n$ values. Therefore, we provide evidence that different combinations of $x_0-n$ can be chosen to engineer intense and localized hotspots within a targeted $z_{hs}$ range on the $\mu$m-scale. 



\textit{Conclusions—}We demonstrate the generation of highly localized intense hotspots using tailored superpositions of finite-energy Airy surface plasmon polaritons at a silver-glass interface. We show how to precisely control their intensity $I^{\rm hs}_{\rm sn}$, size $d_n^{x,y}$ and location $z_{\rm hs}$ by manipulating the ASPP mode order $n$ and spatial separation $x_0$ parameters. The combination of ASPPs is governed by the complex wave number, and the resultant interference pattern leads to hotspots with sub-50-nm sizes and enhancement factors as large as $I^{hs}/I^{single}\approx$ 5. We find that increasing the parameter $x_0$ helps generate nm-sized hotspots as far along the propagation axis as $z=$ 1\,$\mu$m. The combination of ASPP modes $n\leq3$ leads to hotspot enhancement factors well above 3, where higher-order ASPP pairs aid to generate sharper hotspots. The relationship between the location of hotspots along the propagation axis $z$ and $x_0$ follows a predictable parabolic trajectory, providing an analytical framework for designing customizable field enhancements for a wide range of $z_{\rm hs}$ locations. Our results highlight the versatility of the ASPP mode superposition to tailor the size and intensity of hotspots, which could ultimately be used as a new tool to trap and transport matter at the nanoscale using structureless plasmonic tweezers and to develop novel plasmonic circuits.


\textit{Acknowledgements—}This research has been partially supported by Project 2020-T1/IND-19951 (JHR) funded by the ``Programa de Atracci\'on de Talento de la Comunidad Aut\'onoma de Madrid (Modalidad~1)''; Grants No.~PID2021-127781NB-I00 (ASS) and  No.~PID2022-137569NB-C42 (RMH) both funded by MCIN/AEI/10.13039/501100011033 and the ``European Union NextGenerationEU/PRTR'' (RMH); the FASLIGHT Network, Grant No.~RED2022-134391-T (RMH), funded by MCIN/AEI/10.13039/501100011033. The authors thank M. Pando Poblador for her contribution to the numerical calculations at an early stage of this work.

\bibliography{references}

\begin{thebibliography}{43}%
\makeatletter
\providecommand \@ifxundefined [1]{%
 \@ifx{#1\undefined}
}%
\providecommand \@ifnum [1]{%
 \ifnum #1\expandafter \@firstoftwo
 \else \expandafter \@secondoftwo
 \fi
}%
\providecommand \@ifx [1]{%
 \ifx #1\expandafter \@firstoftwo
 \else \expandafter \@secondoftwo
 \fi
}%
\providecommand \natexlab [1]{#1}%
\providecommand \enquote  [1]{``#1''}%
\providecommand \bibnamefont  [1]{#1}%
\providecommand \bibfnamefont [1]{#1}%
\providecommand \citenamefont [1]{#1}%
\providecommand \href@noop [0]{\@secondoftwo}%
\providecommand \href [0]{\begingroup \@sanitize@url \@href}%
\providecommand \@href[1]{\@@startlink{#1}\@@href}%
\providecommand \@@href[1]{\endgroup#1\@@endlink}%
\providecommand \@sanitize@url [0]{\catcode `\\12\catcode `\$12\catcode `\&12\catcode `\#12\catcode `\^12\catcode `\_12\catcode `\%12\relax}%
\providecommand \@@startlink[1]{}%
\providecommand \@@endlink[0]{}%
\providecommand \url  [0]{\begingroup\@sanitize@url \@url }%
\providecommand \@url [1]{\endgroup\@href {#1}{\urlprefix }}%
\providecommand \urlprefix  [0]{URL }%
\providecommand \Eprint [0]{\href }%
\providecommand \doibase [0]{https://doi.org/}%
\providecommand \selectlanguage [0]{\@gobble}%
\providecommand \bibinfo  [0]{\@secondoftwo}%
\providecommand \bibfield  [0]{\@secondoftwo}%
\providecommand \translation [1]{[#1]}%
\providecommand \BibitemOpen [0]{}%
\providecommand \bibitemStop [0]{}%
\providecommand \bibitemNoStop [0]{.\EOS\space}%
\providecommand \EOS [0]{\spacefactor3000\relax}%
\providecommand \BibitemShut  [1]{\csname bibitem#1\endcsname}%
\let\auto@bib@innerbib\@empty
\bibitem [{\citenamefont {Zayats}\ \emph {et~al.}(2005{\natexlab{a}})\citenamefont {Zayats}, \citenamefont {Smolyaninov},\ and\ \citenamefont {Maradudin}}]{maradudin:PhysRep:2005}%
  \BibitemOpen
  \bibfield  {author} {\bibinfo {author} {\bibfnamefont {A.~V.}\ \bibnamefont {Zayats}}, \bibinfo {author} {\bibfnamefont {I.~I.}\ \bibnamefont {Smolyaninov}},\ and\ \bibinfo {author} {\bibfnamefont {A.~A.}\ \bibnamefont {Maradudin}},\ }\bibfield  {title} {\bibinfo {title} {Nano-optics of surface plasmon polaritons},\ }\href {https://doi.org/https://doi.org/10.1016/j.physrep.2004.11.001} {\bibfield  {journal} {\bibinfo  {journal} {Phys. Rep.}\ }\textbf {\bibinfo {volume} {408}},\ \bibinfo {pages} {131} (\bibinfo {year} {2005}{\natexlab{a}})}\BibitemShut {NoStop}%
\bibitem [{\citenamefont {Manjavacas}\ and\ \citenamefont {García~de Abajo}(2009)}]{doi:10.1021/nl802044t}%
  \BibitemOpen
  \bibfield  {author} {\bibinfo {author} {\bibfnamefont {A.}~\bibnamefont {Manjavacas}}\ and\ \bibinfo {author} {\bibfnamefont {F.~J.}\ \bibnamefont {García~de Abajo}},\ }\bibfield  {title} {\bibinfo {title} {Robust plasmon waveguides in strongly interacting nanowire arrays},\ }\href {https://doi.org/10.1021/nl802044t} {\bibfield  {journal} {\bibinfo  {journal} {Nano Lett.}\ }\textbf {\bibinfo {volume} {9}},\ \bibinfo {pages} {1285} (\bibinfo {year} {2009})}\BibitemShut {NoStop}%
\bibitem [{\citenamefont {Manjavacas}\ and\ \citenamefont {de~Abajo}(2009)}]{Manjavacas:09}%
  \BibitemOpen
  \bibfield  {author} {\bibinfo {author} {\bibfnamefont {A.}~\bibnamefont {Manjavacas}}\ and\ \bibinfo {author} {\bibfnamefont {F.~J.~G.}\ \bibnamefont {de~Abajo}},\ }\bibfield  {title} {\bibinfo {title} {Coupling of gap plasmons in multi-wire waveguides},\ }\href {https://doi.org/10.1364/OE.17.019401} {\bibfield  {journal} {\bibinfo  {journal} {Opt. Express}\ }\textbf {\bibinfo {volume} {17}},\ \bibinfo {pages} {19401} (\bibinfo {year} {2009})}\BibitemShut {NoStop}%
\bibitem [{\citenamefont {Zhang}\ \emph {et~al.}(2012)\citenamefont {Zhang}, \citenamefont {Zhang},\ and\ \citenamefont {Xu}}]{Zhang:JPhysD:2012}%
  \BibitemOpen
  \bibfield  {author} {\bibinfo {author} {\bibfnamefont {J.}~\bibnamefont {Zhang}}, \bibinfo {author} {\bibfnamefont {L.}~\bibnamefont {Zhang}},\ and\ \bibinfo {author} {\bibfnamefont {W.}~\bibnamefont {Xu}},\ }\bibfield  {title} {\bibinfo {title} {Surface plasmon polaritons: physics and applications},\ }\href {https://doi.org/10.1088/0022-3727/45/11/113001} {\bibfield  {journal} {\bibinfo  {journal} {J. Phys. D: Appl. Phys.}\ }\textbf {\bibinfo {volume} {45}},\ \bibinfo {pages} {113001} (\bibinfo {year} {2012})}\BibitemShut {NoStop}%
\bibitem [{\citenamefont {Luo}\ and\ \citenamefont {Yan}(2012)}]{xiangang:IEEEPhotJ:2012}%
  \BibitemOpen
  \bibfield  {author} {\bibinfo {author} {\bibfnamefont {X.}~\bibnamefont {Luo}}\ and\ \bibinfo {author} {\bibfnamefont {L.}~\bibnamefont {Yan}},\ }\bibfield  {title} {\bibinfo {title} {Surface plasmon polaritons and its applications},\ }\href {https://doi.org/10.1109/JPHOT.2012.2189436} {\bibfield  {journal} {\bibinfo  {journal} {IEEE Photonics J.}\ }\textbf {\bibinfo {volume} {4}},\ \bibinfo {pages} {590} (\bibinfo {year} {2012})}\BibitemShut {NoStop}%
\bibitem [{\citenamefont {Lin}\ \emph {et~al.}(2012)\citenamefont {Lin}, \citenamefont {Dellinger}, \citenamefont {Genevet}, \citenamefont {Cluzel}, \citenamefont {de~Fornel},\ and\ \citenamefont {Capasso}}]{lin:PRL:2012}%
  \BibitemOpen
  \bibfield  {author} {\bibinfo {author} {\bibfnamefont {J.}~\bibnamefont {Lin}}, \bibinfo {author} {\bibfnamefont {J.}~\bibnamefont {Dellinger}}, \bibinfo {author} {\bibfnamefont {P.}~\bibnamefont {Genevet}}, \bibinfo {author} {\bibfnamefont {B.}~\bibnamefont {Cluzel}}, \bibinfo {author} {\bibfnamefont {F.}~\bibnamefont {de~Fornel}},\ and\ \bibinfo {author} {\bibfnamefont {F.}~\bibnamefont {Capasso}},\ }\bibfield  {title} {\bibinfo {title} {Cosine-gauss plasmon beam: A localized long-range nondiffracting surface wave},\ }\href {https://doi.org/10.1103/PhysRevLett.109.093904} {\bibfield  {journal} {\bibinfo  {journal} {Phys. Rev. Lett.}\ }\textbf {\bibinfo {volume} {109}},\ \bibinfo {pages} {093904} (\bibinfo {year} {2012})}\BibitemShut {NoStop}%
\bibitem [{\citenamefont {Puerto}\ \emph {et~al.}(2016)\citenamefont {Puerto}, \citenamefont {Garcia-Lechuga}, \citenamefont {Hernandez-Rueda}, \citenamefont {Garcia-Leis}, \citenamefont {Sanchez-Cortes}, \citenamefont {Solis},\ and\ \citenamefont {Siegel}}]{puerto2016femtosecond}%
  \BibitemOpen
  \bibfield  {author} {\bibinfo {author} {\bibfnamefont {D.}~\bibnamefont {Puerto}}, \bibinfo {author} {\bibfnamefont {M.}~\bibnamefont {Garcia-Lechuga}}, \bibinfo {author} {\bibfnamefont {J.}~\bibnamefont {Hernandez-Rueda}}, \bibinfo {author} {\bibfnamefont {A.}~\bibnamefont {Garcia-Leis}}, \bibinfo {author} {\bibfnamefont {S.}~\bibnamefont {Sanchez-Cortes}}, \bibinfo {author} {\bibfnamefont {J.}~\bibnamefont {Solis}},\ and\ \bibinfo {author} {\bibfnamefont {J.}~\bibnamefont {Siegel}},\ }\bibfield  {title} {\bibinfo {title} {Femtosecond laser-controlled self-assembly of amorphous-crystalline nanogratings in silicon},\ }\href@noop {} {\bibfield  {journal} {\bibinfo  {journal} {Nanotechnology}\ }\textbf {\bibinfo {volume} {27}},\ \bibinfo {pages} {265602} (\bibinfo {year} {2016})}\BibitemShut {NoStop}%
\bibitem [{\citenamefont {Anwar}\ \emph {et~al.}(2018)\citenamefont {Anwar}, \citenamefont {Ning},\ and\ \citenamefont {Mao}}]{anwar:DigCommunNet:2018}%
  \BibitemOpen
  \bibfield  {author} {\bibinfo {author} {\bibfnamefont {R.~S.}\ \bibnamefont {Anwar}}, \bibinfo {author} {\bibfnamefont {H.}~\bibnamefont {Ning}},\ and\ \bibinfo {author} {\bibfnamefont {L.}~\bibnamefont {Mao}},\ }\bibfield  {title} {\bibinfo {title} {Recent advancements in surface plasmon polaritons-plasmonics in subwavelength structures in microwave and terahertz regimes},\ }\href {https://doi.org/https://doi.org/10.1016/j.dcan.2017.08.004} {\bibfield  {journal} {\bibinfo  {journal} {Digit. Commun. Netw.}\ }\textbf {\bibinfo {volume} {4}},\ \bibinfo {pages} {244} (\bibinfo {year} {2018})}\BibitemShut {NoStop}%
\bibitem [{\citenamefont {Zayats}\ \emph {et~al.}(2005{\natexlab{b}})\citenamefont {Zayats}, \citenamefont {Smolyaninov},\ and\ \citenamefont {Maradudin}}]{zayats2005nano}%
  \BibitemOpen
  \bibfield  {author} {\bibinfo {author} {\bibfnamefont {A.~V.}\ \bibnamefont {Zayats}}, \bibinfo {author} {\bibfnamefont {I.~I.}\ \bibnamefont {Smolyaninov}},\ and\ \bibinfo {author} {\bibfnamefont {A.~A.}\ \bibnamefont {Maradudin}},\ }\bibfield  {title} {\bibinfo {title} {Nano-optics of surface plasmon polaritons},\ }\href@noop {} {\bibfield  {journal} {\bibinfo  {journal} {Phys. Rep.}\ }\textbf {\bibinfo {volume} {408}},\ \bibinfo {pages} {131} (\bibinfo {year} {2005}{\natexlab{b}})}\BibitemShut {NoStop}%
\bibitem [{\citenamefont {Minovich}\ \emph {et~al.}(2011)\citenamefont {Minovich}, \citenamefont {Klein}, \citenamefont {Janunts}, \citenamefont {Pertsch}, \citenamefont {Neshev},\ and\ \citenamefont {Kivshar}}]{minovich2011generation}%
  \BibitemOpen
  \bibfield  {author} {\bibinfo {author} {\bibfnamefont {A.}~\bibnamefont {Minovich}}, \bibinfo {author} {\bibfnamefont {A.~E.}\ \bibnamefont {Klein}}, \bibinfo {author} {\bibfnamefont {N.}~\bibnamefont {Janunts}}, \bibinfo {author} {\bibfnamefont {T.}~\bibnamefont {Pertsch}}, \bibinfo {author} {\bibfnamefont {D.~N.}\ \bibnamefont {Neshev}},\ and\ \bibinfo {author} {\bibfnamefont {Y.~S.}\ \bibnamefont {Kivshar}},\ }\bibfield  {title} {\bibinfo {title} {Generation and near-field imaging of {Airy} surface plasmons},\ }\href@noop {} {\bibfield  {journal} {\bibinfo  {journal} {Phys. Rev. Lett.}\ }\textbf {\bibinfo {volume} {107}},\ \bibinfo {pages} {116802} (\bibinfo {year} {2011})}\BibitemShut {NoStop}%
\bibitem [{\citenamefont {Rotenberg}\ and\ \citenamefont {Kuipers}(2014)}]{rotenberg2014mapping}%
  \BibitemOpen
  \bibfield  {author} {\bibinfo {author} {\bibfnamefont {N.}~\bibnamefont {Rotenberg}}\ and\ \bibinfo {author} {\bibfnamefont {L.}~\bibnamefont {Kuipers}},\ }\bibfield  {title} {\bibinfo {title} {Mapping nanoscale light fields},\ }\href@noop {} {\bibfield  {journal} {\bibinfo  {journal} {Nat. Photonics}\ }\textbf {\bibinfo {volume} {8}},\ \bibinfo {pages} {919} (\bibinfo {year} {2014})}\BibitemShut {NoStop}%
\bibitem [{\citenamefont {Mej{\'\i}a-Salazar}\ and\ \citenamefont {Oliveira~Jr}(2018)}]{mejia2018plasmonic}%
  \BibitemOpen
  \bibfield  {author} {\bibinfo {author} {\bibfnamefont {J.}~\bibnamefont {Mej{\'\i}a-Salazar}}\ and\ \bibinfo {author} {\bibfnamefont {O.~N.}\ \bibnamefont {Oliveira~Jr}},\ }\bibfield  {title} {\bibinfo {title} {Plasmonic biosensing: Focus review},\ }\href@noop {} {\bibfield  {journal} {\bibinfo  {journal} {Chem. Rev.}\ }\textbf {\bibinfo {volume} {118}},\ \bibinfo {pages} {10617} (\bibinfo {year} {2018})}\BibitemShut {NoStop}%
\bibitem [{\citenamefont {Rodrigo}\ \emph {et~al.}(2015)\citenamefont {Rodrigo}, \citenamefont {Limaj}, \citenamefont {Janner}, \citenamefont {Etezadi}, \citenamefont {Garc{\'\i}a~de Abajo}, \citenamefont {Pruneri},\ and\ \citenamefont {Altug}}]{rodrigo2015mid}%
  \BibitemOpen
  \bibfield  {author} {\bibinfo {author} {\bibfnamefont {D.}~\bibnamefont {Rodrigo}}, \bibinfo {author} {\bibfnamefont {O.}~\bibnamefont {Limaj}}, \bibinfo {author} {\bibfnamefont {D.}~\bibnamefont {Janner}}, \bibinfo {author} {\bibfnamefont {D.}~\bibnamefont {Etezadi}}, \bibinfo {author} {\bibfnamefont {F.~J.}\ \bibnamefont {Garc{\'\i}a~de Abajo}}, \bibinfo {author} {\bibfnamefont {V.}~\bibnamefont {Pruneri}},\ and\ \bibinfo {author} {\bibfnamefont {H.}~\bibnamefont {Altug}},\ }\bibfield  {title} {\bibinfo {title} {Mid-infrared plasmonic biosensing with graphene},\ }\href@noop {} {\bibfield  {journal} {\bibinfo  {journal} {Science}\ }\textbf {\bibinfo {volume} {349}},\ \bibinfo {pages} {165} (\bibinfo {year} {2015})}\BibitemShut {NoStop}%
\bibitem [{\citenamefont {Brolo}(2012)}]{brolo2012plasmonics}%
  \BibitemOpen
  \bibfield  {author} {\bibinfo {author} {\bibfnamefont {A.~G.}\ \bibnamefont {Brolo}},\ }\bibfield  {title} {\bibinfo {title} {Plasmonics for future biosensors},\ }\href@noop {} {\bibfield  {journal} {\bibinfo  {journal} {Nat. Photonics}\ }\textbf {\bibinfo {volume} {6}},\ \bibinfo {pages} {709} (\bibinfo {year} {2012})}\BibitemShut {NoStop}%
\bibitem [{\citenamefont {Camden}\ \emph {et~al.}(2008)\citenamefont {Camden}, \citenamefont {Dieringer}, \citenamefont {Zhao},\ and\ \citenamefont {Van~Duyne}}]{camden2008controlled}%
  \BibitemOpen
  \bibfield  {author} {\bibinfo {author} {\bibfnamefont {J.~P.}\ \bibnamefont {Camden}}, \bibinfo {author} {\bibfnamefont {J.~A.}\ \bibnamefont {Dieringer}}, \bibinfo {author} {\bibfnamefont {J.}~\bibnamefont {Zhao}},\ and\ \bibinfo {author} {\bibfnamefont {R.~P.}\ \bibnamefont {Van~Duyne}},\ }\bibfield  {title} {\bibinfo {title} {Controlled plasmonic nanostructures for surface-enhanced spectroscopy and sensing},\ }\href@noop {} {\bibfield  {journal} {\bibinfo  {journal} {Acc. of Chem. Res.}\ }\textbf {\bibinfo {volume} {41}},\ \bibinfo {pages} {1653} (\bibinfo {year} {2008})}\BibitemShut {NoStop}%
\bibitem [{\citenamefont {Jahn}\ \emph {et~al.}(2016)\citenamefont {Jahn}, \citenamefont {Patze}, \citenamefont {Hidi}, \citenamefont {Knipper}, \citenamefont {Radu}, \citenamefont {M{\"u}hlig}, \citenamefont {Y{\"u}ksel}, \citenamefont {Peksa}, \citenamefont {Weber}, \citenamefont {Mayerh{\"o}fer} \emph {et~al.}}]{jahn2016plasmonic}%
  \BibitemOpen
  \bibfield  {author} {\bibinfo {author} {\bibfnamefont {M.}~\bibnamefont {Jahn}}, \bibinfo {author} {\bibfnamefont {S.}~\bibnamefont {Patze}}, \bibinfo {author} {\bibfnamefont {I.~J.}\ \bibnamefont {Hidi}}, \bibinfo {author} {\bibfnamefont {R.}~\bibnamefont {Knipper}}, \bibinfo {author} {\bibfnamefont {A.~I.}\ \bibnamefont {Radu}}, \bibinfo {author} {\bibfnamefont {A.}~\bibnamefont {M{\"u}hlig}}, \bibinfo {author} {\bibfnamefont {S.}~\bibnamefont {Y{\"u}ksel}}, \bibinfo {author} {\bibfnamefont {V.}~\bibnamefont {Peksa}}, \bibinfo {author} {\bibfnamefont {K.}~\bibnamefont {Weber}}, \bibinfo {author} {\bibfnamefont {T.}~\bibnamefont {Mayerh{\"o}fer}}, \emph {et~al.},\ }\bibfield  {title} {\bibinfo {title} {Plasmonic nanostructures for surface enhanced spectroscopic methods},\ }\href@noop {} {\bibfield  {journal} {\bibinfo  {journal} {Analyst}\ }\textbf {\bibinfo {volume} {141}},\ \bibinfo {pages} {756} (\bibinfo {year} {2016})}\BibitemShut {NoStop}%
\bibitem [{\citenamefont {Miroshnichenko}\ and\ \citenamefont {Kivshar}(2013)}]{miroshnichenko2013polarization}%
  \BibitemOpen
  \bibfield  {author} {\bibinfo {author} {\bibfnamefont {A.~E.}\ \bibnamefont {Miroshnichenko}}\ and\ \bibinfo {author} {\bibfnamefont {Y.~S.}\ \bibnamefont {Kivshar}},\ }\bibfield  {title} {\bibinfo {title} {Polarization traffic control for surface plasmons},\ }\href@noop {} {\bibfield  {journal} {\bibinfo  {journal} {Science}\ }\textbf {\bibinfo {volume} {340}},\ \bibinfo {pages} {283} (\bibinfo {year} {2013})}\BibitemShut {NoStop}%
\bibitem [{\citenamefont {Gong}\ \emph {et~al.}(2018)\citenamefont {Gong}, \citenamefont {Alpeggiani}, \citenamefont {Sciacca}, \citenamefont {Garnett},\ and\ \citenamefont {Kuipers}}]{gong2018nanoscale}%
  \BibitemOpen
  \bibfield  {author} {\bibinfo {author} {\bibfnamefont {S.-H.}\ \bibnamefont {Gong}}, \bibinfo {author} {\bibfnamefont {F.}~\bibnamefont {Alpeggiani}}, \bibinfo {author} {\bibfnamefont {B.}~\bibnamefont {Sciacca}}, \bibinfo {author} {\bibfnamefont {E.~C.}\ \bibnamefont {Garnett}},\ and\ \bibinfo {author} {\bibfnamefont {L.}~\bibnamefont {Kuipers}},\ }\bibfield  {title} {\bibinfo {title} {Nanoscale chiral valley-photon interface through optical spin-orbit coupling},\ }\href@noop {} {\bibfield  {journal} {\bibinfo  {journal} {Science}\ }\textbf {\bibinfo {volume} {359}},\ \bibinfo {pages} {443} (\bibinfo {year} {2018})}\BibitemShut {NoStop}%
\bibitem [{\citenamefont {Bliokh}\ \emph {et~al.}(2017)\citenamefont {Bliokh}, \citenamefont {Bekshaev},\ and\ \citenamefont {Nori}}]{bliokh2017optical}%
  \BibitemOpen
  \bibfield  {author} {\bibinfo {author} {\bibfnamefont {K.~Y.}\ \bibnamefont {Bliokh}}, \bibinfo {author} {\bibfnamefont {A.~Y.}\ \bibnamefont {Bekshaev}},\ and\ \bibinfo {author} {\bibfnamefont {F.}~\bibnamefont {Nori}},\ }\bibfield  {title} {\bibinfo {title} {Optical momentum and angular momentum in complex media: from the {Abraham--Minkowski} debate to unusual properties of surface plasmon-polaritons},\ }\href@noop {} {\bibfield  {journal} {\bibinfo  {journal} {New J. of Phys.}\ }\textbf {\bibinfo {volume} {19}},\ \bibinfo {pages} {123014} (\bibinfo {year} {2017})}\BibitemShut {NoStop}%
\bibitem [{\citenamefont {Noordam}\ \emph {et~al.}(2020)\citenamefont {Noordam}, \citenamefont {Hernandez-Rueda}, \citenamefont {Talsma},\ and\ \citenamefont {Kuipers}}]{noordam2020plasmon}%
  \BibitemOpen
  \bibfield  {author} {\bibinfo {author} {\bibfnamefont {M.}~\bibnamefont {Noordam}}, \bibinfo {author} {\bibfnamefont {J.}~\bibnamefont {Hernandez-Rueda}}, \bibinfo {author} {\bibfnamefont {L.}~\bibnamefont {Talsma}},\ and\ \bibinfo {author} {\bibfnamefont {L.}~\bibnamefont {Kuipers}},\ }\bibfield  {title} {\bibinfo {title} {Plasmon-induced enhancement of nonlinear optical processes in a double-resonant metallic nanostructure grating},\ }\href@noop {} {\bibfield  {journal} {\bibinfo  {journal} {Appl. Phys. Lett.}\ }\textbf {\bibinfo {volume} {116}} (\bibinfo {year} {2020})}\BibitemShut {NoStop}%
\bibitem [{\citenamefont {Forbes}\ \emph {et~al.}(2021)\citenamefont {Forbes}, \citenamefont {de~Oliveira},\ and\ \citenamefont {Dennis}}]{forbes:NatPhotonics:2021}%
  \BibitemOpen
  \bibfield  {author} {\bibinfo {author} {\bibfnamefont {A.}~\bibnamefont {Forbes}}, \bibinfo {author} {\bibfnamefont {M.}~\bibnamefont {de~Oliveira}},\ and\ \bibinfo {author} {\bibfnamefont {M.~R.}\ \bibnamefont {Dennis}},\ }\bibfield  {title} {\bibinfo {title} {Structured light},\ }\href {https://doi.org/10.1038/s41566-021-00780-4} {\bibfield  {journal} {\bibinfo  {journal} {Nat. Photonics}\ }\textbf {\bibinfo {volume} {15}},\ \bibinfo {pages} {253} (\bibinfo {year} {2021})}\BibitemShut {NoStop}%
\bibitem [{\citenamefont {Baumgartl}\ \emph {et~al.}(2008)\citenamefont {Baumgartl}, \citenamefont {Mazilu},\ and\ \citenamefont {Dholakia}}]{dholakia:NatPhoton:2008}%
  \BibitemOpen
  \bibfield  {author} {\bibinfo {author} {\bibfnamefont {J.}~\bibnamefont {Baumgartl}}, \bibinfo {author} {\bibfnamefont {M.}~\bibnamefont {Mazilu}},\ and\ \bibinfo {author} {\bibfnamefont {K.}~\bibnamefont {Dholakia}},\ }\bibfield  {title} {\bibinfo {title} {Optically mediated particle clearing using {Airy} wavepackets},\ }\href {https://doi.org/10.1038/nphoton.2008.201} {\bibfield  {journal} {\bibinfo  {journal} {Nat. Photonics}\ }\textbf {\bibinfo {volume} {2}},\ \bibinfo {pages} {675} (\bibinfo {year} {2008})}\BibitemShut {NoStop}%
\bibitem [{\citenamefont {Vettenburg}\ \emph {et~al.}(2014)\citenamefont {Vettenburg}, \citenamefont {Dalgarno}, \citenamefont {Nylk}, \citenamefont {Coll-Llad\'o}, \citenamefont {Ferrier}, \citenamefont {\v{C}i\v{z}m\'ar}, \citenamefont {Gunn-Moore},\ and\ \citenamefont {Dholakia}}]{dholakia:NatMethods:2014}%
  \BibitemOpen
  \bibfield  {author} {\bibinfo {author} {\bibfnamefont {T.}~\bibnamefont {Vettenburg}}, \bibinfo {author} {\bibfnamefont {H.~I.~C.}\ \bibnamefont {Dalgarno}}, \bibinfo {author} {\bibfnamefont {J.}~\bibnamefont {Nylk}}, \bibinfo {author} {\bibfnamefont {C.}~\bibnamefont {Coll-Llad\'o}}, \bibinfo {author} {\bibfnamefont {D.~E.~K.}\ \bibnamefont {Ferrier}}, \bibinfo {author} {\bibfnamefont {T.}~\bibnamefont {\v{C}i\v{z}m\'ar}}, \bibinfo {author} {\bibfnamefont {F.~J.}\ \bibnamefont {Gunn-Moore}},\ and\ \bibinfo {author} {\bibfnamefont {K.}~\bibnamefont {Dholakia}},\ }\bibfield  {title} {\bibinfo {title} {Light-sheet microscopy using an {Airy} beam},\ }\href {https://doi.org/10.1038/nmeth.2922} {\bibfield  {journal} {\bibinfo  {journal} {Nat. Methods}\ }\textbf {\bibinfo {volume} {11}},\ \bibinfo {pages} {541} (\bibinfo {year} {2014})}\BibitemShut {NoStop}%
\bibitem [{\citenamefont {Fang}\ and\ \citenamefont {Sun}(2015)}]{fang:LSA:2015}%
  \BibitemOpen
  \bibfield  {author} {\bibinfo {author} {\bibfnamefont {Y.}~\bibnamefont {Fang}}\ and\ \bibinfo {author} {\bibfnamefont {M.}~\bibnamefont {Sun}},\ }\bibfield  {title} {\bibinfo {title} {Nanoplasmonic waveguides: towards applications in integrated nanophotonic circuits},\ }\href {https://doi.org/10.1038/lsa.2015.67} {\bibfield  {journal} {\bibinfo  {journal} {Light Sci. Appl.}\ }\textbf {\bibinfo {volume} {4}},\ \bibinfo {pages} {e294} (\bibinfo {year} {2015})}\BibitemShut {NoStop}%
\bibitem [{\citenamefont {Liu}\ \emph {et~al.}(2005)\citenamefont {Liu}, \citenamefont {Steele}, \citenamefont {Srituravanich}, \citenamefont {Pikus}, \citenamefont {Sun},\ and\ \citenamefont {Zhang}}]{liu2005focusing}%
  \BibitemOpen
  \bibfield  {author} {\bibinfo {author} {\bibfnamefont {Z.}~\bibnamefont {Liu}}, \bibinfo {author} {\bibfnamefont {J.~M.}\ \bibnamefont {Steele}}, \bibinfo {author} {\bibfnamefont {W.}~\bibnamefont {Srituravanich}}, \bibinfo {author} {\bibfnamefont {Y.}~\bibnamefont {Pikus}}, \bibinfo {author} {\bibfnamefont {C.}~\bibnamefont {Sun}},\ and\ \bibinfo {author} {\bibfnamefont {X.}~\bibnamefont {Zhang}},\ }\bibfield  {title} {\bibinfo {title} {Focusing surface plasmons with a plasmonic lens},\ }\href@noop {} {\bibfield  {journal} {\bibinfo  {journal} {Nano Lett.}\ }\textbf {\bibinfo {volume} {5}},\ \bibinfo {pages} {1726} (\bibinfo {year} {2005})}\BibitemShut {NoStop}%
\bibitem [{\citenamefont {Min}\ \emph {et~al.}(2013)\citenamefont {Min}, \citenamefont {Shen}, \citenamefont {Shen}, \citenamefont {Zhang}, \citenamefont {Fang}, \citenamefont {Yuan}, \citenamefont {Du}, \citenamefont {Zhu}, \citenamefont {Lei},\ and\ \citenamefont {Yuan}}]{min2013focused}%
  \BibitemOpen
  \bibfield  {author} {\bibinfo {author} {\bibfnamefont {C.}~\bibnamefont {Min}}, \bibinfo {author} {\bibfnamefont {Z.}~\bibnamefont {Shen}}, \bibinfo {author} {\bibfnamefont {J.}~\bibnamefont {Shen}}, \bibinfo {author} {\bibfnamefont {Y.}~\bibnamefont {Zhang}}, \bibinfo {author} {\bibfnamefont {H.}~\bibnamefont {Fang}}, \bibinfo {author} {\bibfnamefont {G.}~\bibnamefont {Yuan}}, \bibinfo {author} {\bibfnamefont {L.}~\bibnamefont {Du}}, \bibinfo {author} {\bibfnamefont {S.}~\bibnamefont {Zhu}}, \bibinfo {author} {\bibfnamefont {T.}~\bibnamefont {Lei}},\ and\ \bibinfo {author} {\bibfnamefont {X.}~\bibnamefont {Yuan}},\ }\bibfield  {title} {\bibinfo {title} {Focused plasmonic trapping of metallic particles},\ }\href@noop {} {\bibfield  {journal} {\bibinfo  {journal} {Nat. Communications}\ }\textbf {\bibinfo {volume} {4}},\ \bibinfo {pages} {2891} (\bibinfo {year} {2013})}\BibitemShut {NoStop}%
\bibitem [{\citenamefont {Zhang}\ \emph {et~al.}(2021)\citenamefont {Zhang}, \citenamefont {Min}, \citenamefont {Dou}, \citenamefont {Wang}, \citenamefont {Urbach}, \citenamefont {Somekh},\ and\ \citenamefont {Yuan}}]{zhang2021plasmonic}%
  \BibitemOpen
  \bibfield  {author} {\bibinfo {author} {\bibfnamefont {Y.}~\bibnamefont {Zhang}}, \bibinfo {author} {\bibfnamefont {C.}~\bibnamefont {Min}}, \bibinfo {author} {\bibfnamefont {X.}~\bibnamefont {Dou}}, \bibinfo {author} {\bibfnamefont {X.}~\bibnamefont {Wang}}, \bibinfo {author} {\bibfnamefont {H.~P.}\ \bibnamefont {Urbach}}, \bibinfo {author} {\bibfnamefont {M.~G.}\ \bibnamefont {Somekh}},\ and\ \bibinfo {author} {\bibfnamefont {X.}~\bibnamefont {Yuan}},\ }\bibfield  {title} {\bibinfo {title} {Plasmonic tweezers: for nanoscale optical trapping and beyond},\ }\href@noop {} {\bibfield  {journal} {\bibinfo  {journal} {Light Sci. Appl.}\ }\textbf {\bibinfo {volume} {10}},\ \bibinfo {pages} {59} (\bibinfo {year} {2021})}\BibitemShut {NoStop}%
\bibitem [{\citenamefont {Dreher}\ \emph {et~al.}(2023)\citenamefont {Dreher}, \citenamefont {Janoschka}, \citenamefont {Frank}, \citenamefont {Giessen},\ and\ \citenamefont {Meyer~zu Heringdorf}}]{dreher2023focused}%
  \BibitemOpen
  \bibfield  {author} {\bibinfo {author} {\bibfnamefont {P.}~\bibnamefont {Dreher}}, \bibinfo {author} {\bibfnamefont {D.}~\bibnamefont {Janoschka}}, \bibinfo {author} {\bibfnamefont {B.}~\bibnamefont {Frank}}, \bibinfo {author} {\bibfnamefont {H.}~\bibnamefont {Giessen}},\ and\ \bibinfo {author} {\bibfnamefont {F.-J.}\ \bibnamefont {Meyer~zu Heringdorf}},\ }\bibfield  {title} {\bibinfo {title} {Focused surface plasmon polaritons coherently couple to electronic states in above-threshold electron emission},\ }\href@noop {} {\bibfield  {journal} {\bibinfo  {journal} {Communications Physics}\ }\textbf {\bibinfo {volume} {6}},\ \bibinfo {pages} {15} (\bibinfo {year} {2023})}\BibitemShut {NoStop}%
\bibitem [{\citenamefont {Salandrino}\ and\ \citenamefont {Christodoulides}(2010)}]{christodoulides:OL:2010}%
  \BibitemOpen
  \bibfield  {author} {\bibinfo {author} {\bibfnamefont {A.}~\bibnamefont {Salandrino}}\ and\ \bibinfo {author} {\bibfnamefont {D.~N.}\ \bibnamefont {Christodoulides}},\ }\bibfield  {title} {\bibinfo {title} {Airy plasmon: a nondiffracting surface wave},\ }\href {https://doi.org/10.1364/OL.35.002082} {\bibfield  {journal} {\bibinfo  {journal} {Opt. Lett.}\ }\textbf {\bibinfo {volume} {35}},\ \bibinfo {pages} {2082} (\bibinfo {year} {2010})}\BibitemShut {NoStop}%
\bibitem [{\citenamefont {Li}\ \emph {et~al.}(2011)\citenamefont {Li}, \citenamefont {Li}, \citenamefont {Wang}, \citenamefont {Zhang},\ and\ \citenamefont {Zhu}}]{li:PRL:2011}%
  \BibitemOpen
  \bibfield  {author} {\bibinfo {author} {\bibfnamefont {L.}~\bibnamefont {Li}}, \bibinfo {author} {\bibfnamefont {T.}~\bibnamefont {Li}}, \bibinfo {author} {\bibfnamefont {S.~M.}\ \bibnamefont {Wang}}, \bibinfo {author} {\bibfnamefont {C.}~\bibnamefont {Zhang}},\ and\ \bibinfo {author} {\bibfnamefont {S.~N.}\ \bibnamefont {Zhu}},\ }\bibfield  {title} {\bibinfo {title} {Plasmonic {Airy} beam generated by in-plane diffraction},\ }\href {https://doi.org/10.1103/PhysRevLett.107.126804} {\bibfield  {journal} {\bibinfo  {journal} {Phys. Rev. Lett.}\ }\textbf {\bibinfo {volume} {107}},\ \bibinfo {pages} {126804} (\bibinfo {year} {2011})}\BibitemShut {NoStop}%
\bibitem [{\citenamefont {Klein}\ \emph {et~al.}(2012)\citenamefont {Klein}, \citenamefont {Minovich}, \citenamefont {Steinert}, \citenamefont {Janunts}, \citenamefont {T\"{u}nnermann}, \citenamefont {Neshev}, \citenamefont {Kivshar},\ and\ \citenamefont {Pertsch}}]{Klein:OptLett:2012}%
  \BibitemOpen
  \bibfield  {author} {\bibinfo {author} {\bibfnamefont {A.~E.}\ \bibnamefont {Klein}}, \bibinfo {author} {\bibfnamefont {A.}~\bibnamefont {Minovich}}, \bibinfo {author} {\bibfnamefont {M.}~\bibnamefont {Steinert}}, \bibinfo {author} {\bibfnamefont {N.}~\bibnamefont {Janunts}}, \bibinfo {author} {\bibfnamefont {A.}~\bibnamefont {T\"{u}nnermann}}, \bibinfo {author} {\bibfnamefont {D.~N.}\ \bibnamefont {Neshev}}, \bibinfo {author} {\bibfnamefont {Y.~S.}\ \bibnamefont {Kivshar}},\ and\ \bibinfo {author} {\bibfnamefont {T.}~\bibnamefont {Pertsch}},\ }\bibfield  {title} {\bibinfo {title} {Controlling plasmonic hot spots by interfering {Airy} beams},\ }\href {https://doi.org/10.1364/OL.37.003402} {\bibfield  {journal} {\bibinfo  {journal} {Opt. Lett.}\ }\textbf {\bibinfo {volume} {37}},\ \bibinfo {pages} {3402} (\bibinfo {year} {2012})}\BibitemShut {NoStop}%
\bibitem [{\citenamefont {Minovich}\ \emph {et~al.}(2014)\citenamefont {Minovich}, \citenamefont {Klein}, \citenamefont {Neshev}, \citenamefont {Pertsch}, \citenamefont {Kivshar},\ and\ \citenamefont {Christodoulides}}]{minovich:LasPhotRev:2014}%
  \BibitemOpen
  \bibfield  {author} {\bibinfo {author} {\bibfnamefont {A.~E.}\ \bibnamefont {Minovich}}, \bibinfo {author} {\bibfnamefont {A.~E.}\ \bibnamefont {Klein}}, \bibinfo {author} {\bibfnamefont {D.~N.}\ \bibnamefont {Neshev}}, \bibinfo {author} {\bibfnamefont {T.}~\bibnamefont {Pertsch}}, \bibinfo {author} {\bibfnamefont {Y.~S.}\ \bibnamefont {Kivshar}},\ and\ \bibinfo {author} {\bibfnamefont {D.~N.}\ \bibnamefont {Christodoulides}},\ }\bibfield  {title} {\bibinfo {title} {{Airy} plasmons: non-diffracting optical surface waves},\ }\href {https://doi.org/https://doi.org/10.1002/lpor.201300055} {\bibfield  {journal} {\bibinfo  {journal} {Laser Photonics Rev.}\ }\textbf {\bibinfo {volume} {8}},\ \bibinfo {pages} {221} (\bibinfo {year} {2014})}\BibitemShut {NoStop}%
\bibitem [{\citenamefont {Pitarke}\ \emph {et~al.}(2006)\citenamefont {Pitarke}, \citenamefont {Silkin}, \citenamefont {Chulkov},\ and\ \citenamefont {Echenique}}]{echenique:RepProgPhys:2006}%
  \BibitemOpen
  \bibfield  {author} {\bibinfo {author} {\bibfnamefont {J.~M.}\ \bibnamefont {Pitarke}}, \bibinfo {author} {\bibfnamefont {V.~M.}\ \bibnamefont {Silkin}}, \bibinfo {author} {\bibfnamefont {E.~V.}\ \bibnamefont {Chulkov}},\ and\ \bibinfo {author} {\bibfnamefont {P.~M.}\ \bibnamefont {Echenique}},\ }\bibfield  {title} {\bibinfo {title} {Theory of surface plasmons and surface-plasmon polaritons},\ }\href {https://doi.org/10.1088/0034-4885/70/1/R01} {\bibfield  {journal} {\bibinfo  {journal} {Rep. Prog. Phys.}\ }\textbf {\bibinfo {volume} {70}},\ \bibinfo {pages} {1} (\bibinfo {year} {2006})}\BibitemShut {NoStop}%
\bibitem [{\citenamefont {Mart{\'\i}nez-Herrero}\ and\ \citenamefont {Manjavacas}(2016)}]{manjavacas:PRA:2016}%
  \BibitemOpen
  \bibfield  {author} {\bibinfo {author} {\bibfnamefont {R.}~\bibnamefont {Mart{\'\i}nez-Herrero}}\ and\ \bibinfo {author} {\bibfnamefont {A.}~\bibnamefont {Manjavacas}},\ }\bibfield  {title} {\bibinfo {title} {Basis for paraxial surface-plasmon-polariton packets},\ }\href {https://doi.org/10.1103/PhysRevA.94.063829} {\bibfield  {journal} {\bibinfo  {journal} {Phys. Rev. A}\ }\textbf {\bibinfo {volume} {94}},\ \bibinfo {pages} {063829} (\bibinfo {year} {2016})}\BibitemShut {NoStop}%
\bibitem [{\citenamefont {Hu}\ \emph {et~al.}(2021)\citenamefont {Hu}, \citenamefont {Xu}, \citenamefont {Lin},\ and\ \citenamefont {Deng}}]{hu:ResPhys:2021}%
  \BibitemOpen
  \bibfield  {author} {\bibinfo {author} {\bibfnamefont {H.}~\bibnamefont {Hu}}, \bibinfo {author} {\bibfnamefont {C.}~\bibnamefont {Xu}}, \bibinfo {author} {\bibfnamefont {M.}~\bibnamefont {Lin}},\ and\ \bibinfo {author} {\bibfnamefont {D.}~\bibnamefont {Deng}},\ }\bibfield  {title} {\bibinfo {title} {{Pearcey} plasmon: An autofocusing surface wave},\ }\href {https://doi.org/https://doi.org/10.1016/j.rinp.2021.104416} {\bibfield  {journal} {\bibinfo  {journal} {Res. Phys.}\ }\textbf {\bibinfo {volume} {26}},\ \bibinfo {pages} {104416} (\bibinfo {year} {2021})}\BibitemShut {NoStop}%
\bibitem [{\citenamefont {Ruan}\ \emph {et~al.}(2024)\citenamefont {Ruan}, \citenamefont {Zhang}, \citenamefont {Chen}, \citenamefont {Feng}, \citenamefont {Li}, \citenamefont {Wu}, \citenamefont {Wen}, \citenamefont {Wang},\ and\ \citenamefont {Deng}}]{ruan:OL:2024}%
  \BibitemOpen
  \bibfield  {author} {\bibinfo {author} {\bibfnamefont {Z.}~\bibnamefont {Ruan}}, \bibinfo {author} {\bibfnamefont {J.}~\bibnamefont {Zhang}}, \bibinfo {author} {\bibfnamefont {Y.}~\bibnamefont {Chen}}, \bibinfo {author} {\bibfnamefont {Z.}~\bibnamefont {Feng}}, \bibinfo {author} {\bibfnamefont {Y.}~\bibnamefont {Li}}, \bibinfo {author} {\bibfnamefont {H.}~\bibnamefont {Wu}}, \bibinfo {author} {\bibfnamefont {S.}~\bibnamefont {Wen}}, \bibinfo {author} {\bibfnamefont {G.}~\bibnamefont {Wang}},\ and\ \bibinfo {author} {\bibfnamefont {D.}~\bibnamefont {Deng}},\ }\bibfield  {title} {\bibinfo {title} {{Pearcey Talbot}-like plasmon: a plasmonic bottle array generation scheme},\ }\href {https://doi.org/10.1364/OL.531141} {\bibfield  {journal} {\bibinfo  {journal} {Opt. Lett.}\ }\textbf {\bibinfo {volume} {49}},\ \bibinfo {pages} {4673} (\bibinfo {year} {2024})}\BibitemShut {NoStop}%
\bibitem [{\citenamefont {Chen}\ \emph {et~al.}(2023)\citenamefont {Chen}, \citenamefont {Tu}, \citenamefont {Hu}, \citenamefont {Zhang}, \citenamefont {Feng}, \citenamefont {Wang}, \citenamefont {Hong},\ and\ \citenamefont {Deng}}]{chen:OL:2023}%
  \BibitemOpen
  \bibfield  {author} {\bibinfo {author} {\bibfnamefont {Y.}~\bibnamefont {Chen}}, \bibinfo {author} {\bibfnamefont {Z.}~\bibnamefont {Tu}}, \bibinfo {author} {\bibfnamefont {H.}~\bibnamefont {Hu}}, \bibinfo {author} {\bibfnamefont {J.}~\bibnamefont {Zhang}}, \bibinfo {author} {\bibfnamefont {Z.}~\bibnamefont {Feng}}, \bibinfo {author} {\bibfnamefont {Z.}~\bibnamefont {Wang}}, \bibinfo {author} {\bibfnamefont {W.}~\bibnamefont {Hong}},\ and\ \bibinfo {author} {\bibfnamefont {D.}~\bibnamefont {Deng}},\ }\bibfield  {title} {\bibinfo {title} {Olver plasmon: an accelerating surface wave with various orders},\ }\href {https://doi.org/10.1364/OL.487750} {\bibfield  {journal} {\bibinfo  {journal} {Opt. Lett.}\ }\textbf {\bibinfo {volume} {48}},\ \bibinfo {pages} {2030} (\bibinfo {year} {2023})}\BibitemShut {NoStop}%
\bibitem [{\citenamefont {Hernandez-Rueda}\ \emph {et~al.}(2024)\citenamefont {Hernandez-Rueda}, \citenamefont {Sanz},\ and\ \citenamefont {Mart{\'\i}nez-Herrero}}]{hernandez2024engineering}%
  \BibitemOpen
  \bibfield  {author} {\bibinfo {author} {\bibfnamefont {J.}~\bibnamefont {Hernandez-Rueda}}, \bibinfo {author} {\bibfnamefont {A.~S.}\ \bibnamefont {Sanz}},\ and\ \bibinfo {author} {\bibfnamefont {R.}~\bibnamefont {Mart{\'\i}nez-Herrero}},\ }\bibfield  {title} {\bibinfo {title} {Engineering of self-bending surface plasmon polaritons through {Hermite-Gaussian} mode expansion},\ }\href@noop {} {\bibfield  {journal} {\bibinfo  {journal} {arXiv preprint arXiv:2412.08706}\ } (\bibinfo {year} {2024})}\BibitemShut {NoStop}%
\bibitem [{\citenamefont {Siviloglou}\ and\ \citenamefont {Christodoulides}(2007)}]{christodoulides:OptLett:2007}%
  \BibitemOpen
  \bibfield  {author} {\bibinfo {author} {\bibfnamefont {G.~A.}\ \bibnamefont {Siviloglou}}\ and\ \bibinfo {author} {\bibfnamefont {D.~N.}\ \bibnamefont {Christodoulides}},\ }\bibfield  {title} {\bibinfo {title} {Accelerating finite energy {Airy} beams},\ }\href {https://doi.org/10.1364/OL.32.000979} {\bibfield  {journal} {\bibinfo  {journal} {Opt. Lett.}\ }\textbf {\bibinfo {volume} {32}},\ \bibinfo {pages} {979} (\bibinfo {year} {2007})}\BibitemShut {NoStop}%
\bibitem [{\citenamefont {Johnson}\ and\ \citenamefont {Christy}(1972)}]{johnson:PRB:1972}%
  \BibitemOpen
  \bibfield  {author} {\bibinfo {author} {\bibfnamefont {P.~B.}\ \bibnamefont {Johnson}}\ and\ \bibinfo {author} {\bibfnamefont {R.~W.}\ \bibnamefont {Christy}},\ }\bibfield  {title} {\bibinfo {title} {Optical constants of the noble metals},\ }\href {https://doi.org/10.1103/PhysRevB.6.4370} {\bibfield  {journal} {\bibinfo  {journal} {Phys. Rev. B}\ }\textbf {\bibinfo {volume} {6}},\ \bibinfo {pages} {4370} (\bibinfo {year} {1972})}\BibitemShut {NoStop}%
\bibitem [{\citenamefont {Kotsifaki}\ \emph {et~al.}(2020)\citenamefont {Kotsifaki}, \citenamefont {Truong},\ and\ \citenamefont {Chormaic}}]{kotsifaki2020fano}%
  \BibitemOpen
  \bibfield  {author} {\bibinfo {author} {\bibfnamefont {D.~G.}\ \bibnamefont {Kotsifaki}}, \bibinfo {author} {\bibfnamefont {V.~G.}\ \bibnamefont {Truong}},\ and\ \bibinfo {author} {\bibfnamefont {S.~N.}\ \bibnamefont {Chormaic}},\ }\bibfield  {title} {\bibinfo {title} {Fano-resonant, asymmetric, metamaterial-assisted tweezers for single nanoparticle trapping},\ }\href@noop {} {\bibfield  {journal} {\bibinfo  {journal} {Nano Lett.}\ }\textbf {\bibinfo {volume} {20}},\ \bibinfo {pages} {3388} (\bibinfo {year} {2020})}\BibitemShut {NoStop}%
\bibitem [{\citenamefont {Kotsifaki}\ \emph {et~al.}(2021)\citenamefont {Kotsifaki}, \citenamefont {Truong},\ and\ \citenamefont {Nic~Chormaic}}]{kotsifaki2021dynamic}%
  \BibitemOpen
  \bibfield  {author} {\bibinfo {author} {\bibfnamefont {D.~G.}\ \bibnamefont {Kotsifaki}}, \bibinfo {author} {\bibfnamefont {V.~G.}\ \bibnamefont {Truong}},\ and\ \bibinfo {author} {\bibfnamefont {S.}~\bibnamefont {Nic~Chormaic}},\ }\bibfield  {title} {\bibinfo {title} {Dynamic multiple nanoparticle trapping using metamaterial plasmonic tweezers},\ }\href@noop {} {\bibfield  {journal} {\bibinfo  {journal} {Appl. Phys. Lett.}\ }\textbf {\bibinfo {volume} {118}} (\bibinfo {year} {2021})}\BibitemShut {NoStop}%
\bibitem [{\citenamefont {Kotsifaki}\ \emph {et~al.}(2024)\citenamefont {Kotsifaki}, \citenamefont {Truong}, \citenamefont {Dindo}, \citenamefont {Laurino},\ and\ \citenamefont {Chormaic}}]{kotsifaki2024hybrid}%
  \BibitemOpen
  \bibfield  {author} {\bibinfo {author} {\bibfnamefont {D.~G.}\ \bibnamefont {Kotsifaki}}, \bibinfo {author} {\bibfnamefont {V.~G.}\ \bibnamefont {Truong}}, \bibinfo {author} {\bibfnamefont {M.}~\bibnamefont {Dindo}}, \bibinfo {author} {\bibfnamefont {P.}~\bibnamefont {Laurino}},\ and\ \bibinfo {author} {\bibfnamefont {S.~N.}\ \bibnamefont {Chormaic}},\ }\bibfield  {title} {\bibinfo {title} {Hybrid metamaterial optical tweezers for dielectric particles and biomolecules discrimination},\ }\href@noop {} {\bibfield  {journal} {\bibinfo  {journal} {arXiv preprint arXiv:2402.12878}\ } (\bibinfo {year} {2024})}\BibitemShut {NoStop}%
\end{thebibliography}%


%

\appendix

\textit{Appendix: Finite-energy self-bending SPPs—}In the following, we provide a summary of the procedure to formally describe the propagation of Gaussian SPPs considering the case of paraxial Airy self-bending SPP in terms of the Hermite-Gaussian modes used in the main text. An in-depth derivation of the procedure can be found in \cite{hernandez2024engineering}. A general expression of a finite-energy self-bending SPP at a dielectric-metal interface ($x-z$ plane) in the paraxial approximation is given by

\begin{equation}
 \psi_0(x,z) = \int e^{i\varphi(u)} \xi(u) e^{i|k_{\rm sp}|ux} e^{-ik^*_{\rm sp} u^2z/2} du ,
 \label{eq8}
\end{equation}
where $\varphi(u)$ is a real-valued polynomial that determines the type of self-bending SPP beam, which in the present case is an Airy plasmon, with $\varphi(u) = u^3/3$. The SPP finite energy condition can be defined as
\begin{equation}
 \int |\xi(u)|^2 du < \infty .
 \label{eq9}
\end{equation}

We recast $\xi(u)$ as a linear superposition of Hermite-Gaussian modes [complete set of orthonormal functions $\Psi_n(u)$ described in Eq.~\eqref{eq5}], as
\begin{equation}
 \xi(u) = \sum c_n \Psi_n(u) ,
 \label{eq10}
\end{equation}
with the coefficients given by
\begin{equation}
 c_n = \int \Psi_n (u) \xi (u) du .
 \label{eq11}
\end{equation}
Finally, we can obtain a mode expansion of the field amplitude by combining Eqs.~\eqref{eq10} and~\eqref{eq8} as
\begin{equation}
 \psi_0(x,z) = \sum c_n \Psi_{n0} (x,z) ,
 \label{eq26}
\end{equation}
where each mode $n$ is given by
\begin{equation}
 \Psi_{n0}(x,z) = \int e^{iu^3/3} \Psi_n(u) e^{i|k_{\rm sp}|xu - ik^*_{\rm sp} u^2z/2} du .
 \label{eq27}
\end{equation}
Note here that the coefficients $c_n$ depend on $\alpha$ through $\Psi_n(u)$, thus each specific choice of $\alpha$ defines a family of modes.


\end{document}